\documentclass[a4paper,conference]{IEEEtran}
\IEEEoverridecommandlockouts

\usepackage{cite}
\usepackage{amsmath,amssymb,amsfonts}
\usepackage{algorithmic}
\usepackage{graphicx}
\usepackage{listings}
\usepackage{textcomp}
\usepackage{xcolor}
\usepackage{balance}
\usepackage{booktabs}
\usepackage{tikz}
\usepackage{pgfplots}
\usepackage{tcolorbox}
\usepackage{adjustbox}
\usepackage{tabularx}
\usepackage{threeparttable}
\usepackage{multirow}
\usepackage{url}
\usepackage{hyperref}
\usepackage{colortbl}
\usepackage[capitalize,nameinlink,noabbrev]{cleveref}
\crefname{figure}{Fig.}{Figs.}

\usepgfplotslibrary{statistics}

\def\BibTeX{{\rm B\kern-.05em{\sc i\kern-.025em b}\kern-.08em
    T\kern-.1667em\lower.7ex\hbox{E}\kern-.125emX}}

\begin{document}

\newcommand{\RqOne}{RQ$_1$: What kind of code snippet descriptions do developers write in READMEs?}

\newcommand{\RqTwo}{RQ$_2$: How accurately can LLMs classify different types of code snippet descriptions?}

\newcommand{\RqThree}{RQ$_3$: Can LLM assist developers in creating code snippet descriptions?}

\title{Uncovering Intention through LLM-Driven Code Snippet Description Generation}



\makeatletter
\author{
\begin{@IEEEauthorhalign}
\IEEEauthorblockN{\hspace*{-0.2cm}1\textsuperscript{st} Yusuf Sulistyo Nugroho}
\IEEEauthorblockA{\hspace*{-0.2cm}\textit{Informatics Engineering} \\
\hspace*{-0.2cm}\textit{Universitas Muhammadiyah Surakarta}\\
\hspace*{-0.2cm}Surakarta, Indonesia \\
\hspace*{-0.2cm}yusuf.nugroho@ums.ac.id}
\\ [1.5ex]
\IEEEauthorblockN{\hspace*{-0.2cm}4\textsuperscript{th} Raula Gaikovina Kula}
\IEEEauthorblockA{\hspace*{-0.2cm}\textit{Graduate School of IST} \\
\hspace*{-0.2cm}\textit{University of Osaka}\\
\hspace*{-0.2cm}Osaka, Japan \\
\hspace*{-0.2cm}raula-k@ist.osaka-u.ac.jp}
\and
\IEEEauthorblockN{2\textsuperscript{nd} Farah Danisha Salam}
\IEEEauthorblockA{\textit{Informatics Engineering} \\
\textit{Universitas Muhammadiyah Surakarta}\\
Surakarta, Indonesia \\
L200204017@student.ums.ac.id}
\\ [1.5ex]
\IEEEauthorblockN{5\textsuperscript{th} Kazumasa Shimari}
\IEEEauthorblockA{\textit{Information Science} \\
\textit{Nara Institute of Science and Technology}\\
Nara, Japan \\
k.shimari@is.naist.jp}
\and
\IEEEauthorblockN{\hspace*{0.2cm}3\textsuperscript{rd} Brittany Reid}
\IEEEauthorblockA{\hspace*{0.2cm}\textit{Information Science} \\
\hspace*{0.2cm}\textit{Nara Institute of Science and Technology}\\
\hspace*{0.2cm}Nara, Japan \\
\hspace*{0.5cm}brittany.reid@naist.ac.jp} 
\\ [1.5ex]
\IEEEauthorblockN{\hspace*{0.2cm}6\textsuperscript{th} Kenichi Matsumoto}
\IEEEauthorblockA{\hspace*{0.2cm}\textit{Information Science} \\
\hspace*{0.2cm}\textit{Nara Institute of Science and Technology}\\
\hspace*{0.2cm}Nara, Japan \\
\hspace*{0.2cm}matumoto@is.naist.jp}
\end{@IEEEauthorhalign}
}
\makeatother

\maketitle

\begin{abstract}
Documenting code snippets is essential to pinpoint key areas where both developers and users should pay attention. Examples include usage examples and other Application Programming Interfaces (APIs), which are especially important for third-party libraries. With the rise of Large Language Models (LLMs), the key goal is to investigate the kinds of description developers commonly use and evaluate how well an LLM, in this case Llama, can support description generation. We use NPM Code Snippets, consisting of 185,412 packages with 1,024,579 code snippets. From there, we use 400 code snippets (and their descriptions) as samples. 
First, our manual classification found that the majority of original descriptions (55.5\%) highlight example-based usage.
This finding emphasizes the importance of clear documentation, as some descriptions lacked sufficient detail to convey intent.
Second, the LLM correctly identified the majority of original descriptions as ``Example'' (79.75\%), which is identical to our manual finding, showing a propensity for generalization. 
Third, compared to the originals, the produced description had an average similarity score of 0.7173, suggesting relevance but room for improvement. Scores below 0.9 indicate some irrelevance. 
Our results show that depending on the task of the code snippet, the intention of the document may differ from being instructions for usage, installations, or descriptive learning examples for any user of a library.
\end{abstract}


\section{Introduction}

In software development, the README file is an important document at the earliest stages, providing essential information about the project to users and/or other developers. Good documentation can help the stakeholders to understand, utilize, maintain, and evolve the related system~\cite{aghajani2020software}.
One important element often found in README files is the code snippet, which showcases the main functionality of the software. 
Each code snippet typically includes a description that explains its purpose or functionality, thereby improving code readability and maintainability~\cite{7886920}. 

\begin{figure}[]
    \centering
    \includegraphics[width=0.5\textwidth]{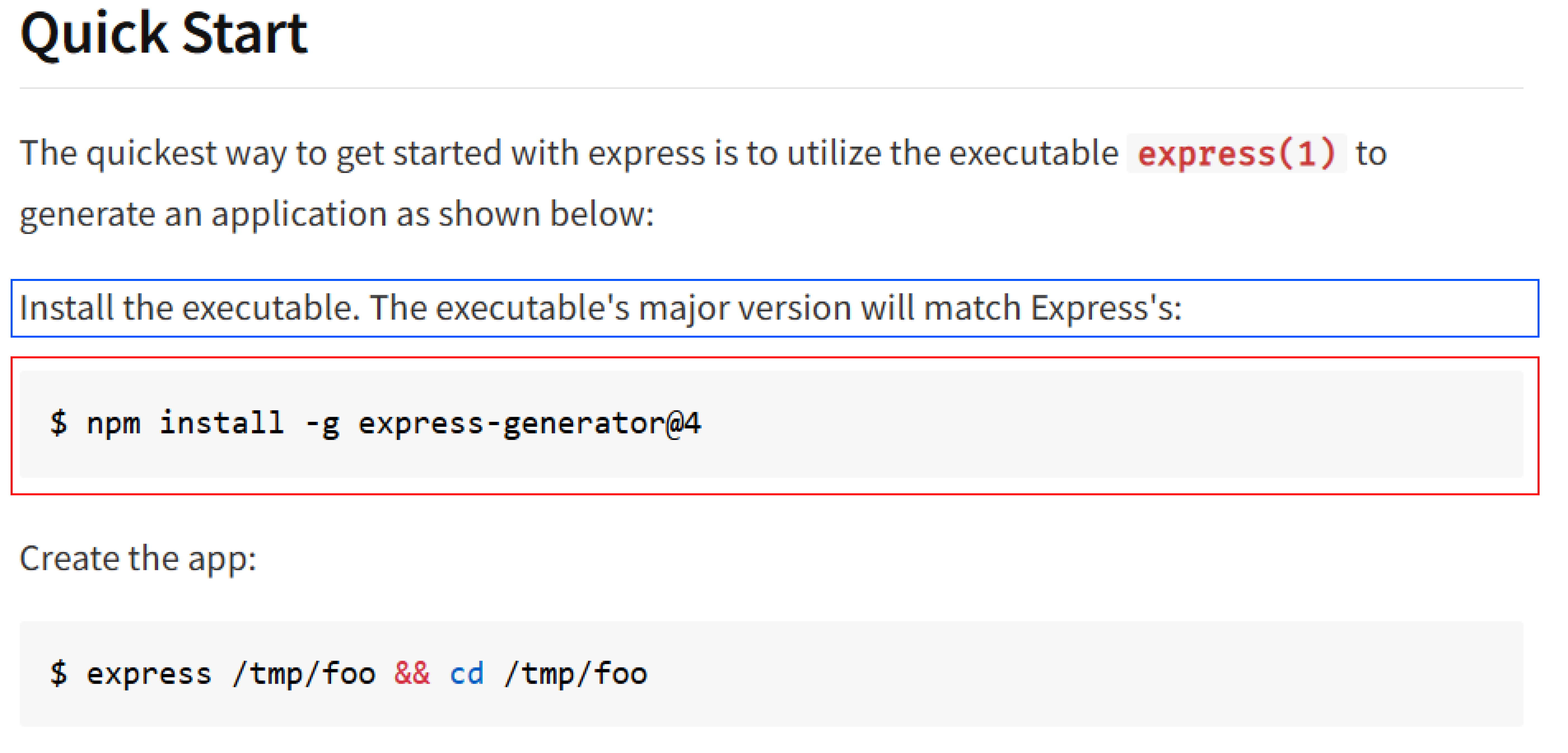}
    \caption{Example of code snippet with description taken from npm express.}
    \label{fig:ReadMeexample}
\end{figure}

Recent advances in documentation automation have led to the integration of LLMs in generating and classifying technical texts. Tools powered by LLMs are now being explored for various software engineering tasks, such as code summarization~\cite{yun2024project}, documentation generation~\cite{huang2024let}, software testing~\cite{wang2024software}, and automated bug reproduction~\cite{kang2024evaluating}. These developments set the stage for exploring how LLMs can assist with specific documentation elements, such as README-embedded code snippet descriptions.

Developers often include code snippets in their README files to explain `what' or `how' the repository functions~\cite{prana2019categorizing,liu2022readme}.
\cref{fig:ReadMeexample} presents an example of a code snippet from an NPM package, where the blue box contains a description and the red box contains the code. 
Understanding the nature of these code snippet descriptions and how they are written has become increasingly important, particularly with the advancement of automation tools. 
LLMs have demonstrated impressive capabilities across a range of tasks such as natural language processing~\cite{brown2020language}, education~\cite{kasneci2023chatgpt}, healthcare~\cite{daungsupawong2024innovative}, and code generation~\cite{bappon2024autogenics}. However, applying LLMs to code documentation tasks like classifying or generating code snippet descriptions in README files introduces challenges such as hallucinations, ambiguity in intent, and alignment with human expectations~\cite{jiang2024surveylargelanguagemodels,fang2024largelanguagemodelscode}.

Previous studies have analyzed README content to understand developers' documentation practices. For example, Prana et al.~\cite{prana2019categorizing} proposed a classifier to automatically categorize sections in README files into categories like ``Why,'' ``How,'' and ``What,'' while Ikeda et al.~\cite{ikeda2019empirical} examined the structure of JavaScript project READMEs that include instructions for installation, usage, and donation guidelines. Other work has investigated README files as indicators of software maturity~\cite{xia2023understanding} or synergy between software projects~\cite{el2021quantifying}. Although these studies focus on the broader content structure of README files, to the best of our knowledge, there is limited exploration of the specific role and classification of code snippet descriptions.

In this study, we focus on evaluating the effectiveness of LLMs in distinguishing between types of code snippet descriptions and comparing LLM-generated descriptions with the original. The objective focuses on assessing the accuracy and readability of LLM-produced descriptions and determining whether LLMs can serve as helpful tools for developers in writing code snippet descriptions.
To achieve this, we utilized 400 code snippet descriptions extracted from prior study~\cite{10050780} for both manual and automatic classifications using predefined categories and prompts.

To guide our investigation, we pose the following three research questions and their brief results: 
\begin{itemize}
    \item \RqOne

    Our manual analysis in RQ$_1$ revealed that developers predominantly use example-based descriptions, followed by instructional ones.

    \item \RqTwo

    In RQ$_2$, we found that the LLM similarly favoured the example category but showed a stronger bias toward it, suggesting alignment with but also a divergence from human classification.

    \item \RqThree

    For RQ$_3$, the LLM-generated descriptions achieved moderate similarity with the original ones, indicating potential support for documentation tasks but highlighting the need for developer oversight.
\end{itemize}





\section{Related Work}

Many prior studies have investigated the role of LLMs in code documentation and summarization. For example, research has examined how LLMs, such as GPT-3 and LLaMA, can be leveraged for generating inline code comments, summarizing software repositories, and enhancing API documentation. MacNeil et al.~\cite{macneil2022generating} investigated the ability of GPT-3 to generate diverse and contextually accurate explanations for code snippets, revealing that although LLMs can provide useful summaries, they often suffer from verbosity or hallucination. Similarly, Huang et al.~\cite{huang2023bias} conducted a study on bias testing and mitigation in LLM-based code generation, highlighting the unfairness in code generation, including the risk of undesirable and malicious software behaviors. 

In the context of software documentation, a recent study by Koreeda et al.~\cite{koreeda2023larch} utilized LLMs in automating README content generation, with findings suggesting that the proposed tool can generate coherent and factually correct READMEs in the majority of cases.
Prana et al.~\cite{prana2019categorizing} conducted a qualitative study on 4,226 README sections from 393 GitHub repositories and developed a multi-label classifier that categorizes README content with an F1 score of 0.746. Their findings highlight that while information on the repository’s functionality is common, many READMEs lack details on the purpose and project status, and automated section labeling improves information discovery for developers. Meanwhile, Ikeda et al.~\cite{ikeda2019empirical} examined JavaScript README contents, providing insights into common documentation patterns. 
Although these works focus on broader README content generation and categorization, they do not explore the code snippet descriptions specifically. To the best of our knowledge, our study is the first to systematically categorize and generate README code snippet descriptions using LLMs, bridging the gap between manual classification and AI-assisted documentation.

\section{Research Methods}
In this section, we describe the data collection and the design of prompts to input into the LLM. In detail, our methodology follows the following three key steps:

\begin{itemize}
    \item Dataset Construction: Sample 400 code snippets with descriptions from a large-scale NPM dataset.

    \item Classification Task: Conduct manual annotation followed by LLM-based classification using a template prompt.

    \item Description Generation and Evaluation: Use the LLM to generate new descriptions, then evaluate their similarity with original descriptions using BERTScore.
\end{itemize}

\begin{figure*}[]
    \centering
    \begin{minipage}{\textwidth}
    \footnotesize
    \begin{lstlisting}[basicstyle=\ttfamily, frame=single]
"""
You are a developer who creates a README file. You have to follow the rules below:
- RESPOND to the answer using the given TYPES.
- Do not INCLUDE any comment.
- MAKE SURE to follow the OUTPUT FORMAT.
- DESCRIPTION should be short and only one line.
- If you can't choose the TYPE or don't understand the code, you can respond with "Couldn't decide a task 
  or description".
- DO NOT answer MORE THAN ONE answer for each question.
- DO NOT IMPROVE or CHANGE the FORMAT, FOLLOW THE OUTPUT FORMAT STRICTLY.

There are three TYPES of code snippet's description:
- Instruction has two types here, those are Installation Instruction and Usage Instruction:
  1. Installation instruction: guide to install and configure software or tools on a computer, including 
     download steps, system requirements, installation, configuration, and verification.
  2. Usage instruction: how to use the installed tool, application, or software in an efficient manner, 
     covers the main commands, features, and fundamental functions.
- Example has three types, those are Usage Example, Feature Explanation, and Code Example:
  1. Usage example: A particular example showing how a program or application's feature or function can be 
     applied in a practical setting.
  2. Feature explanation: a thorough description of a particular feature, component, or attribute of the 
     code snippet.
  3. Code example: A code example usually shows a function or a specific piece of code to only represent 
     the code snippet.
- Unclear, which indicates confusion or lack of clarity.

OUTPUT FORMAT:
Type: Instruction
Option: Installation instruction
Example: guide to install and configure software or tools on a computer, including download steps, system 
requirements, installation, configuration, and verification.
"""
    \end{lstlisting}
    \end{minipage}
    \caption{Example of prompts input to LLM for README file description creation.} 
    \label{fig:promptexample}
\end{figure*}

\subsection{Data Collection}
\label{sec:dataset}

\begin{table}[b]
    \centering
    \caption{Overview of The Dataset Used in This Study~\cite{10050780}}
    \begin{tabular}{|l|r|}
        \hline
        \multicolumn{1}{c|}{\textbf{Field}} & \multicolumn{1}{c|}{\textbf{Value}} \\
        \hline
        Dataset name & NPM Code Snippets \\
        Total number of packages & 185,412 \\
        Total number of code snippets & 1,024,579 \\
        Number of samples used & 400 \\
        \hline
    \end{tabular}
    \label{tab:dataset_info_horizontal}
\end{table}

We outline the steps of our data preparation process for description and code snippets. In this work, we used the NPM code snippets and package information dataset from prior work~\cite{10050780}. 
To obtain the package names, code snippets, and their descriptions, we constructed a dataset that is detailed in~\autoref{tab:dataset_info_horizontal}, consisting of 185,412 packages and 1,024,579 code snippets, each accompanied by at least one textual description. This dataset was selected for its extensive coverage of real-world npm projects, ensuring a diverse and representative sample of documentation styles.

From this dataset, we then extracted the code snippets and their descriptions for analysis. Since we performed a manual classification in this qualitative study, we created a random sample of code snippets taken from the main dataset. Using a survey calculator to define the sample size,\footnote{\url{https://www.surveymonkey.com/mp/sample-size-calculator/}} we applied the confidence level at 95\% and margin of error at 5\%. This calculation yielded 385 samples. However, in this study, we rounded up the size to 400 samples.

\subsection{Prompt Design}
\label{sec:promptDesign}

Prior work has shown the importance of prompt design on the output of LLMs~\cite{reynolds2021prompt}. 
Thus, as presented in~\cref{fig:promptexample}, our structured prompt was designed to provide the LLM with sufficient context to generate meaningful and accurate classifications. A longer prompt was used to mitigate ambiguity and provide clear classification rules, preventing misinterpretation of the task.
Although longer prompts may introduce attention-related issues, we found that structuring them explicitly helped maintain response consistency.
In this study, we use Ollama with the llama3 model to generate the output. The template outlines essential rules for the LLM to follow, minimizing ambiguity and increasing the consistency of its output.

We also define the types of classifications that the LLM is expected to perform, specifying the categories of description and the criteria for each. To illustrate our expectations, we also provide example results that serve as benchmarks for the LLM's responses, as shown in~\cref{fig:promptexample}. This comprehensive approach is crucial for guiding the LLM in generating accurate and relevant insights for our study.

\section{Results and Discussions}
\label{sec:Results}

\subsection{\RqOne}

\begin{table}[]
    \centering
    \caption{Categories of Code Snippets Description Used in Our Analysis}
    \label{tab:descCategories}
    \resizebox{\columnwidth}{!}{
    \begin{tabular}{|l|p{3.5cm}|p{2.5cm}|}
        \hline
        \textbf{Category} & \textbf{Definition} & \textbf{Example} \\
        \hline
        Instruction & If the description describes the instructions for software installation, configuration, or guidance of a tool or application utilization. & \textit{Import everything from a reducer file and create a reducer function from it.} \\
        Example & The description usually provides information about a specific example of applying a feature in practice, a detailed description of a particular feature, or presents code snippet of a function or concept. & \textit{Thus, target, specs and replace Props are available. example:}   \\
        Unclear & If the description indicates confusion or lack of clarity. & \textit{Is equivalent} \\
        \hline
    \end{tabular}
    }
\end{table}

In this analysis, we initially discussed the coding guide that will be applied for the manual classification of the code snippet descriptions. As described in~\autoref{tab:descCategories}, three categories of code snippets are defined for our annotation, which are `instruction', `example', and `unclear.'

To facilitate our manual analysis, we utilized a random sample of 400 code snippet descriptions from the NPM dataset described in Section~\ref{sec:dataset}. To ensure reliability in the manual classifications, we followed a prior study~\cite{nugroho2021project}, where the first two authors independently categorized the first 30 data samples. To reach a consensus, we discussed the disagreement regarding the labeling. We then counted Cohen's Kappa to measure the level of agreement.\footnote{\url{http://justusrandolph.net/kappa/}} 
From the calculation, we achieved the Kappa score at 80\%, which indicates `almost perfect agreement'~\cite{viera2005understanding}. Based on this motivating score, the manual labeling for the remaining samples was then performed individually by a single author.

\begin{table}[]
\centering
\caption{Manual Classification Results}
\begin{tabular}{|l|rr|}
\hline
\multicolumn{1}{|c|}{}                                    & \multicolumn{2}{c|}{\textbf{Number of description}}                                  \\ \cline{2-3} 
\multicolumn{1}{|c|}{\multirow{-2}{*}{\textbf{Category}}} & \multicolumn{1}{c|}{\textbf{Qty}}                & \multicolumn{1}{c|}{\textbf{\%}} \\ \hline
\rowcolor[HTML]{C0C0C0} 
a. Instructions                                                   & \multicolumn{1}{r|}{\cellcolor[HTML]{C0C0C0}176} & 44.00\%                          \\ \hline
- Installation instruction                                & \multicolumn{1}{r|}{14}                          & 3.50\%                           \\ \hline
- Usage instruction                                       & \multicolumn{1}{r|}{162}                         & 40.50\%                          \\ \hline
\rowcolor[HTML]{C0C0C0} 
b. Example                                            & \multicolumn{1}{r|}{\cellcolor[HTML]{C0C0C0}222} & 55.50\%                          \\ \hline
- Usage example                                           & \multicolumn{1}{r|}{38}                          & 9.50\%                           \\ \hline
- Feature explanation                                     & \multicolumn{1}{r|}{157}                         & 39.25\%                          \\ \hline
- Code example                                            & \multicolumn{1}{r|}{27}                          & 6.75\%                           \\ \hline
\rowcolor[HTML]{C0C0C0} 
c. Unclear                                                & \multicolumn{1}{r|}{\cellcolor[HTML]{C0C0C0}2}   & 0.50\%                           \\ \hline
Total                                                     & \multicolumn{1}{r|}{400}                         & 100.00\%                         \\ \hline
\end{tabular}
\label{table2manualclassification}
\end{table}

From the manual classification of the 400 code snippet descriptions, as presented in~\autoref{table2manualclassification}, we found that developers most often write descriptions as `example', with 55.5\% of the samples explaining what the code is about. 
This type of description typically provides context or clarification about the purpose and functionality of the code, helping readers understand its intended use and behavior.
Although it is lower than an example, an `instruction' type of description is also frequently written by developers to indicate instructions for other developers, accounting for 44\% of the samples.

This finding highlights that, in general, software developers are common in providing both instruction and example descriptions.
They give more priority to the explanations of code functionality and instructions for tasks over cautionary notes or future reminders.
In sum, understanding this kind of pattern provides insight into developers' focus during the process of documentation, which can improve the clarity and usability of their code.

\subsection{\RqTwo}

In this experiment, we followed a similar technique as in RQ$_1$ by utilizing the same data samples of code snippet descriptions from our dataset. However, the classification is performed automatically by an LLM rather than manually and uses the template or prompt as LLM's rule, described in Section~\ref{sec:promptDesign}.

\begin{table}[htbp]
\centering
\caption{LLM Classification Results}
\begin{tabular}{|l|rr|}
\hline
\multicolumn{1}{|c|}{}                                    & \multicolumn{2}{c|}{\textbf{Number of description}}                                  \\ \cline{2-3} 
\multicolumn{1}{|c|}{\multirow{-2}{*}{\textbf{Category}}} & \multicolumn{1}{c|}{\textbf{Qty}}                & \multicolumn{1}{c|}{\textbf{\%}} \\ \hline
\rowcolor[HTML]{C0C0C0} 
a. Instruction                                                   & \multicolumn{1}{r|}{\cellcolor[HTML]{C0C0C0}81} & 20.25\%                          \\ \hline
- Installation instruction                                & \multicolumn{1}{r|}{55}                          & 13.75\%                           \\ \hline
- Usage instruction                                       & \multicolumn{1}{r|}{26}                         & 6.50\%                          \\ \hline
\rowcolor[HTML]{C0C0C0} 
b. Example                                            & \multicolumn{1}{r|}{\cellcolor[HTML]{C0C0C0}319} & 79.75\%                          \\ \hline
- Usage example                                           & \multicolumn{1}{r|}{36}                          & 9.00\%                           \\ \hline
- Feature explanation                                     & \multicolumn{1}{r|}{140}                         & 35.00\%                          \\ \hline
- Code example                                            & \multicolumn{1}{r|}{143}                          & 35.75\%                           \\ \hline
\rowcolor[HTML]{C0C0C0} 
c. Unclear                                                & \multicolumn{1}{r|}{\cellcolor[HTML]{C0C0C0}0}   & 0.00\%                           \\ \hline
Total                                                     & \multicolumn{1}{r|}{400}                         & 100.00\%                         \\ \hline
\end{tabular}
\label{tabl3LLM classification results}
\end{table}

\autoref{tabl3LLM classification results} describes that the LLM also predominantly categorized the given code snippet descriptions as `example', accounting for 79.75\% of the samples, followed by `instruction' at 20.25\%. 
Although the classification results show a noticeable difference in proportions, they align with our manual findings in RQ$_1$, where `example' constituted the majority of descriptions. Notably, our manual classification was conducted independently by the authors rather than by external developers to ensure a standardized approach.
Despite following the same classification template, LLMs consistently favored the `Example' category over `Instruction.' This discrepancy may arise due to the inherent bias of LLMs toward common patterns in training data, potentially leading to overgeneralization. In addition, LLMs might produce convincing but inaccurate responses, requiring further analysis.

\subsection{\RqThree}

To answer this research question, we used the same 400 descriptions as in RQ$_1$. The LLM, in this experiment, is responsible for creating a new description for a particular code snippet. To evaluate the similarity between the description generated by the LLM and the original description written by developers, we used the BERT (Bidirectional Encoder Representations from Transformers) model~\cite{malik2024tarzan}. By measuring this similarity, we aim to determine whether the LLM can assist developers in creating insightful descriptions for code snippets based only on the code itself.

\begin{figure}[htbp]
    \centering
    \includegraphics[height=.4\linewidth, width=0.49\textwidth]{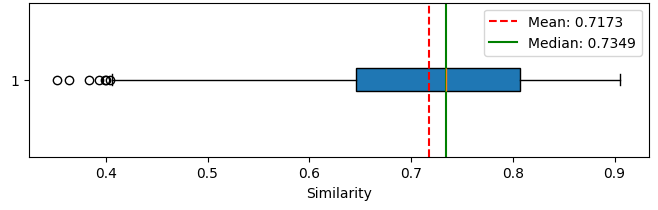}
    \caption{Similarity values between the original and generated comments of code snippets.}
    \label{fig:BERTvalue}
\end{figure}

As presented in~\cref{fig:BERTvalue}, from 400 code snippet descriptions, the average similarity score is 0.72, indicating a relatively high level of similarity between the LLM-generated descriptions and the original ones.
We found that there are many examples of descriptions having high scores of similarity, which are used as the basis for determining the description category. 
Here is one example of a comment with the highest similarity score at 0.958, where both the original and generated descriptions produce similar wording and are classified as an `instruction', as follows.

\begin{itemize}
    \item \textbf{Original description:}
    \textit{Loading and using Audio. To load audio files, you need to use the resource-loader built into PIXI.}
    
    \item \textbf{Generated description:} 
    \textit{How to load and play an audio file using PIXI's loader, including adding the file to the loader, loading it, and playing it with the audio manager.}
\end{itemize}

At the same time, the LLM also generates high similarity scores while classifying the description types differently.
The generated description received a similarity score of 0.8993, yet it is classified differently from the original. The original description is classified as an `instruction', while the generated description is categorized as an `example' according to the LLM's classification.
For example:

\begin{itemize}
    \item \textbf{Original description:}
    \textit{There are 2 ways of creating a simple node: 1 - You can use the createNode factory function.}
    
    \item \textbf{Generated description:} 
    \textit{This code snippet explains how to create a node using the createNode function from the `layout-tree' library, with options for setting the node's name and other properties.}
\end{itemize}

This finding suggests that a similarity score under 0.9 indicates a difference in classification types between the original and generated descriptions. 
Although the LLM-generated descriptions show moderate similarity to the original ones, discrepancies in classification highlight potential challenges. The model often emphasizes commonly occurring terminology, occasionally misclassifying task-oriented descriptions as examples. Future work should explore prompt refinements and dataset adjustments to mitigate this issue.

\section{Threats to Validity}
In this paper, we realize that there are limitations that might threaten the validity of the results.
One important limitation of LLM-generated descriptions is the potential for data leakage. 
Since the LLM may have seen some snippets during training, this could impact the objectivity of results. To mitigate this, we selected samples randomly from a large and diverse dataset and cross-referenced against known training corpora when available. We also emphasized analysis of classification behavior over content memorization.
The structure and phrasing of prompts can also influence the output of LLMs. To reduce bias, we adopted a highly structured and rule-based prompt design that clearly defined the expected classification types and provided output examples. The prompt was kept consistent across all samples.

In addition, the weakness of this work is also related to evaluation. Although full evaluation metrics (e.g., accuracy, F1-score) and human judgment were not implemented, however, we included indirect quantitative signals (such as category distribution alignment and BERT-based similarity scoring) as proxy indicators. We also manually reviewed outlier examples to validate interpretations and clarify limitations within the Results section.

Another concern is hallucination, where the LLM produces plausible yet incorrect classifications or descriptions. For instance, our results show that the LLM rarely classified snippets as ``Unclear,'' suggesting an overconfidence bias. Investigating how different prompt structures influence hallucination frequency remains an open challenge.

\section{Challenges and Future Outlook}
Based on our early results, we now highlight the challenges, research directions, and potential recommendations.

Our study has highlighted the importance of clarity in code snippet descriptions, with 55.5\% of the manual classifications falling under the ``Example'' category. However, two instances of unclear descriptions underscore the need for greater focus on readability in these explanations. The effectiveness of LLMs in classifying code snippet descriptions is also noteworthy, though the tendency to overgeneralize remains.

Interestingly, the LLM successfully recognized all instances without ambiguity, which highlights the capability of the model in classifying the code snippet descriptions.
Although the LLM classified all instances without ambiguity, real-world README files often contain vague or poorly written descriptions. 
Thus, future work could investigate whether LLMs are capable of identifying and flagging ambiguous or low-quality descriptions and explore techniques to calibrate confidence or uncertainty in LLM outputs.
Moreover, improving the prompt design, fine-tuning models on various datasets, and incorporating multimodal context could increase the accuracy and reliability of LLM-driven description classification. 

The study has demonstrated that LLMs can be a useful tool for generating new descriptions based on original code snippets, with an average similarity score of 0.7173 between the original and generated descriptions. However, scores below 0.9 indicate some level of irrelevance in the generated description. This suggests that while LLMs can provide suggestions for writing descriptions, prompts may need to be more specific to achieve greater relevance. 
We are using generalized models, so maybe more specific models that deal with source code might have better results. Also, better prompt engineering and awareness of the situation of the code example might deliver better results.

Our findings highlight that while LLMs can assist in generating and categorizing code snippet descriptions, their outputs require careful validation. The tendency to misclassify instructional descriptions as examples suggests that future prompt refinements and dataset adjustments are necessary. Furthermore, addressing potential data leakage and hallucinations will be crucial for improving the reliability of LLM-based documentation tools.

The collaboration between NLP and software engineering communities is also important to develop more effective tools for code snippet description, as evidenced by the success of this research. Moreover, incorporating human feedback into evaluation frameworks can help ensure that LLMs provide accurate and relevant guidance, reducing the risk of misinterpretation or misapplication.
We highlight that through our experiments, we find that code snippets have either the intention of being instructional, examples, or other needs.
Understanding this difference will be key to the intention of the documentation generated.



\balance


\end{document}